\documentclass[traditabstract,letter]{aa}
\usepackage{natbib}
\usepackage{graphicx}
\usepackage{url}

\begin{document}
\title{\vskip-\baselineskip The column density towards LMC\,X-1}
\titlerunning{The column density towards LMC\,X-1}
\author{
 \mbox{Manfred Hanke\inst{1}} \and
 \mbox{J\"orn Wilms\inst{1}} \and
 \mbox{Michael A. Nowak\inst{2}} \and
 \mbox{Laura Barrag\'an\inst{1}} \and
 \mbox{Norbert S. Schulz\inst{2}}
}
\authorrunning{M.\,Hanke et al.}
\institute{
 Dr. Karl Remeis-Observatory \& ECAP, University of Erlangen-Nuremberg, Sternwartstr.~7, 96049~Bamberg, Germany\\
 e-mail: \texttt{Manfred.Hanke@sternwarte.uni-erlangen.de}
\and
 MIT Kavli Institute for Astrophysics and Space Research, NE80-6077, 77 Mass. Ave., Cambridge, MA~02139, USA
}
\date{Received: 3 November 2009 / Accepted: 17 December 2009}
\abstract{
 We measure the neutral absorption towards the black hole X-ray binary system \mbox{LMC\,\,X-1}
 from six archival soft \mbox{X-ray} spectra
 obtained with the gratings and/or CCD detectors on \textsl{Chandra}, \textsl{XMM-Newton}, and \textsl{Swift}.
 Four spectral models for the soft continuum have been investigated.
 While the \texttt{powerlaw} model may overestimate $N_\mathrm{H}$ considerably,
 the others give consistent results.
 Taking the lower metalicity of the Large Magellanic Cloud into account,
 we find equivalent hydrogen column densities of $N_{\rm H}$=(1.0--1.3)${\times}10^{22}\rm\,cm^{-2}$,
 with a systematic dependence on the orbital phase.
 This variation of the neutral absorption can nearly explain the orbital modulation
 of the soft X-ray flux recently detected with the All Sky Monitor (ASM) on the \textsl{Rossi X-ray Timing Explorer} (\textsl{RXTE}).
}

\maketitle

\section{Introduction}\label{sec:intro}
The extragalactic X-ray sources in the Large Magellanic Cloud (LMC),
our $\approx$48\,kpc distant neighbor galaxy,
were discovered in the late 1960s \citep{Mark1969,Price1971}.
Due to the high density of stars, their optical identifications
were uncertain for a long time.
The X-ray binary system (XRB) \object{\mbox{LMC\,\,X-1}}
is located ${\approx}0\fdg5$ south-southeast of the 30 Doradus star-formation region,
in the NGC~2078 (LMC~N159F) nebula.
\citet{Pakull1980}, \citet{Hutchings1983,Hutchings1987}, and \citet{Cowley1995}
were able to identify the counterpart of \mbox{LMC\,\,X-1}
with a $m_{\rm V}{=}14\fm5$ O7/8 giant \citep{Hutchings1983,NegueruelaCoe2002}.
This has allowed the placement of strong dynamical constraints on the compact object's mass.
\citet{Orosz2009} recently have used optical spectra of this star
-- labeled as `star~\#32' by \citet{Cowley1978} and also often called Pakull's star --
to confirm the black hole (BH) candidacy of \mbox{LMC\,\,X-1}.
They derive an orbital period of 3.909\,d
which is consistent with the modulation of the soft X-ray flux of \mbox{LMC\,\,X-1} \citep{LevineCorbet2006}.
Deriving an extinction of $A_V=2.28\pm0.06$ -- much more than previously assumed --
from the $V{-}K$ color excess,
\citet{Orosz2009} infer
a BH mass of $10.9\pm1.6\,M_\odot$.

The persistent XRB \mbox{LMC\,\,X-1} is
the only dynamically confirmed BH candidate
which so far has only been found in the high/soft (thermal dominant) X-ray spectral state.
That is, its X-ray spectrum can be described by
a multi-temperature disk blackbody component
plus a weak soft ($\Gamma{\gg}2$) power law component
\citep{Ebisawa1989,Schlegel1994,Wilms2001,Nowak2001,Haardt2001,Cui2002,Yao2005}.
In comparison, \mbox{LMC\,X-3} usually shows a similarly soft X-ray spectrum,
but also (partial) transitions to the low/hard state \citep{Wilms2001,SmithDawsonSwank2007},
while \mbox{Cyg\,X-1} regularly transits between the low/hard and a soft-intermediate state
and never reaches the thermal dominant state \citep{Wilms2006}.
\mbox{LMC\,\,X-1} is therefore an ideal target to measure the BH spin parameter $a_*$
from the soft X-ray continuum produced by the relativistic accretion disk.
\citet{Gierlinski2001} constrained $a_*$ to be less than 0.998
from a 24\,ks 0.7--10\,keV \textsl{ASCA}-SIS spectrum.
\citet{Gou2009} have recently reported $a_*{=}0.90^{+0.04}_{-0.09}$
from 18 selected \textsl{RXTE}-PCA spectra with exposures between 5--11\,ks and covering 2.5--20\,keV.
The latter authors fix the column density for the photoelectric absorption
to $N_\mathrm{H}{=}4.6\!\times\!10^{21}\,\rm cm^{-2}$
as reported by \citet{Cui2002} based on low statistics data.

An appropriate description of the absorption is, however,
indispensable for modeling the soft X-ray continuum
and likewise for modeling the visual extinction,
and thus the derivation of the system parameters from the dereddened optical spectrum of the companion star.
In this Letter, we therefore aim to accurately describe the column density towards \mbox{LMC\,\,X-1}.
We describe the data in \S\ref{sec:obs}
and present the methods and our analysis in \S\ref{sec:analysis}.
We summarize and discuss our results in \S\ref{sec:discuss}.

\section{Observations and Data Reduction}\label{sec:obs}
\begin{table*}
 \caption{Log of recent soft X-ray observations of LMC\,X-1 with good S/N (Instruments not considered here are in parenthesis.)}
 \label{tab:obs}
 {\centering\begin{tabular}{ccccccccc}
  \hline\hline
  Obs. & Start Date & Start Date & Exposure & ${\phi_\mathrm{orb}^{(\rm T3)}}^\dagger$ & ${\phi_{\rm orb}^{(\rm F9)}}^\ddagger$ & Satellite & ObsID & Instruments \\
       &            & (MJD)      & (ks)     \\
  \hline
  C1   & 2000-01-16 & 51559.2    & 19       & 0.45--0.51 & 0.50--0.56 & \textsl{Chandra} & 93          & HETGS                   \\
  X1   & 2000-10-21 & 51838.7    & 5--7     & 0.94--0.96 & 0.98--0.01 & \textsl{XMM}     & 0112900101  & PN, RGS\,1+2, (MOS\,1+2)\\
  X2   & 2002-09-26 & 52543.2    & 35       & 0.17--0.28 & 0.21--0.32 & \textsl{XMM}     & 0023940401  & RGS\,1+2, (MOS\,1+2)    \\
  S1   & 2007-10-31 & 54404.7    & 2.4      & 0.37--0.42 & 0.37--0.43 & \textsl{Swift}   & 00037079001 & (BAT), XRT/PC, (UVOT)   \\
  S2   & 2007-12-06 & 54440.4    & 9.8      & 0.49--0.61 & 0.50--0.62 & \textsl{Swift}   & 00037079002 & (BAT), XRT/WT, (UVOT)   \\
  S3   & 2007-12-10 & 54444.1    & 4.4      & 0.43--0.50 & 0.43--0.50 & \textsl{Swift}   & 00037079003 & (BAT), XRT/WT, (UVOT)   \\
  \hline
 \end{tabular}\\}
 \mbox{$^\dagger$ orbital phase calculated from the ephemeris of \citet[Table~3]{Orosz2009}: $T_0={\rm MJD}\;53390.8436$, $P=3.90917$\,d}\\
 \mbox{$^\ddagger$ orbital phase calculated from the ephemeris of \citet[Fig.~9]{Orosz2009}: $T_0={\rm MJD}\;53390.75174$, $P=3.9094$\,d}
\end{table*}
\begin{figure}\centering
 \includegraphics[width=\columnwidth]{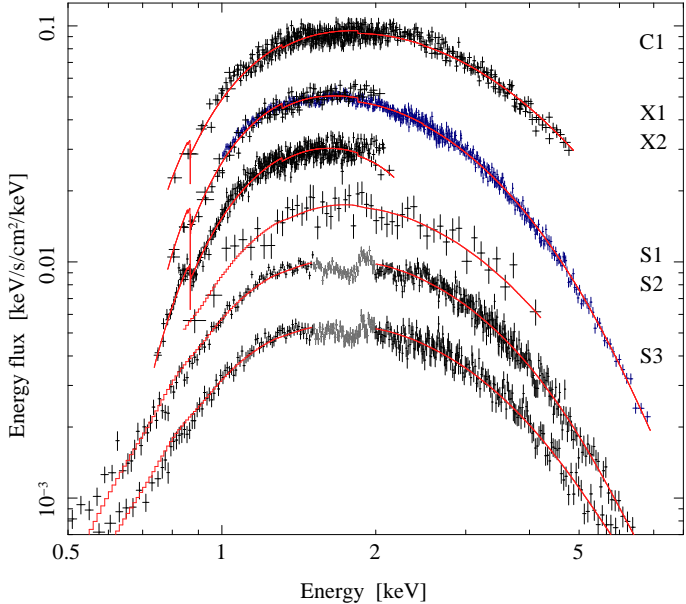}
 \caption{%
  Flux-corrected spectra of \mbox{LMC\,\,X-1} from the six observations,
  shifted in flux according to the labels with respect to C1 for visual clarity.
  The gray data have been ignored because of calibration issues.
  Note that the models shown here for illustrative purposes
  are also broadened by the instrumental response.
 }
 \label{fig:fluxspectra}
 \vskip-1.15\baselineskip
\end{figure}
We study the spectra from all six recent observations with instruments
providing soft X-ray spectra (Table~\ref{tab:obs}).

The \textsl{Chandra} observation C1 was performed using the HETGS \citep{Canizares2005}
and with the detector CCDs operated in timed exposure mode.
The $\pm$first order HEG and MEG spectra as well as the corresponding response matrices
were taken from the \textsl{Chandra} Transmission Grating Catalog archive TGCat\footnote{%
 See \url{http://tgcat.mit.edu}\,.
}.

All instruments of \textsl{XMM-Newton} \citep{Jansen2001} were active during the first (shorter)
\textsl{XMM} observation X1.
The EPIC-pn camera \citep{Strueder2001} was operated in timing mode.
Its data are therefore not affected by photon pile-up \citep{Wilms2003_MQW4}.
The same is true for data from the Reflection Grating Spectrometers \citep[RGS;][]{denHerder2001}
due to their dispersion of the photons,
but not for data from the MOS cameras \citep{Turner2001},
which were operated in full frame imaging mode.
For this reason, we only use the EPIC-pn spectrum
and the first and second order spectra of RGS 1 and 2.
For the second (longer) \textsl{XMM} observation X2,
however, no pn-data are available.
The data were reduced with the Science Analysis Software, \textsc{xmmsas}, v.~7.1,
following standard procedures, i.e.,
applying the SAS tasks \texttt{epchain}, \texttt{emchain}, \texttt{rgsproc},
\texttt{evselect}, \texttt{rmfgen} and \texttt{arfgen}
to produce spectra and response matrices.

\textsl{Swift}'s X-ray telescope \citep[XRT;][]{Burrows2005}
was operated in photon counting (PC) mode during the first \textsl{Swift} observation S1,
which resulted in pile-up.
For S2 and S3, the windowed timing (WT) mode was used.
After reprocessing the data to apply the newest calibration,
spectra were extracted using standard \textsc{ftools}, handled via \texttt{xselect}.
For the PC mode observation S1, we extract an annulus to exclude the region affected by pile-up,
yielding a low quality spectrum only.
Ancillary response files were created with \texttt{xrtmkarf},
and suitable response files for each observation were obtained from the \textsc{caldb}.
The WT mode spectra are not as well calibrated as the PC mode one
around the Si edge (Fig.~\ref{fig:fluxspectra});
we therefore exclude their 1.5--2\,keV data.

All spectral analysis was performed with the
Interactive Spectral Interpretation System \citep[\textsf{ISIS};][]{Houck2000,Noble2006,NobleNowak2008}\footnote{%
 See \url{http://space.mit.edu/cxc/isis/}\,.
}.

\begin{table}
 \caption{Comparison of elemental abundances (by number)
          in~the Galactic ISM and in the LMC as
          $\epsilon(X)=12+\log_{10}(X/{\rm H})$
	 }
 \label{tab:LMCabundances}
 {\centering
 \begin{tabular}{cccc}
  \hline\hline
  \raisebox{-1.2ex}{\rule{0pt}{4ex}}$X$ & $\epsilon_{\rm gal}(X)^{(1)}$ & $\epsilon_{\rm LMC}(X)$ & $10^{\Delta{\textstyle\epsilon}(X)}$ \\
  \hline\raisebox{0cm}[1ex][0ex]{\rule{0pt}{4ex}}
  He & 10.99 & 10.93$^{(5)}$ & 0.87 \\
  C  &  8.38 &  8.03$^{(2)}$ & 0.45 \\
  N  &  7.88 &  7.01$^{(2)}$ & 0.13 \\
  O  &  8.69 &  8.38$^{(2)}$ & 0.49 \\
  Ne &  7.94 &  7.6$^{(4)}$  & 0.46 \\
  Mg &  7.40 &  7.12$^{(2)}$ & 0.53 \\
  Si &  7.27 &  7.21$^{(2)}$ & 0.87 \\
  S  &  7.09 &  6.7$^{(4)}$  & 0.41 \\
  Ar &  6.41 &  6.2$^{(4)}$  & 0.62 \\
  Fe &  7.43 &  7.2$^{(3)}$  & 0.59 \\
  \hline
 \end{tabular}\\}
 \textbf{References}.\\
 (1) \citet{Wilms2000} or using \texttt{xspec\_abund("wilm");} in \textsf{ISIS}.\\
 (2) Przybilla (priv. comm.): average of 7 B-stars in the LMC \citep[see also][]{Korn2002, Korn2005}.\\
 (3) Przybilla (priv. comm.): 1 star in the LMC \citep[see also][]{Przybilla2008}.\\
 (4) \citet{Garnett1999}: H\,\textsc{II} regions in the LMC.\\
 (5) \citet{Dufour1984}.\\
 \textbf{Note}.\\
 The last column is the LMC abundance relative to the Galactic abundance,
 which is a parameter of the \verb|tbvarabs| absorption model \citep[][2009, in prep.]{Wilms2000}.
 For all other elements
 (which hardly contribute to the absorption in the soft X-ray band),
 the average value $10^{\Delta{\textstyle\epsilon}(X)} = 0.5$ is assumed.
 \vskip-.75\baselineskip
\end{table}
\begin{table*}
 \caption{Column density in units of $10^{22}\rm\,cm^{-2}$ for the six observations and the sine fit,  obtained with different continuum models}
 \label{tab:NH}
 {\centering
  \begin{tabular}{cccccccc}
  \hline
  \hline
  \raisebox{-1.2ex}{\rule{0pt}{4ex}} Observation or Fit               & X2                     & S1                     & S3                     & C1                        & S2               & X1                        & Sine Fit \\
  \raisebox{-1.2ex}{\rule{0pt}{4ex}} $\phi_\mathrm{orb}^{\rm (T3)}$             & 0.17--0.28             & 0.37--0.42             & 0.43--0.50             & 0.45--0.51                & 0.49--0.61       & 0.94--0.96                & full orbit \\
  \hline
  \raisebox{-1.2ex}{\rule{0pt}{4ex}} \texttt{diskbb\,+\,powerlaw}$^*$ & $\left(2.00^{+0.17}_{-0.19}\right)^*$ & $\left(1.2^{+0.5}_{-0.2}\right)^*$    & $0.96^{+0.03}_{-0.02}$ & $\left(1.25^{+0.04}_{-0.01}\right)^*$    & $1.031\pm0.017$  & $\left(1.81^{+0.06}_{-0.05}\right)^*$    & $\left(1.43\pm0.43\right)^*$ \\
  \raisebox{-1.2ex}{\rule{0pt}{4ex}} \texttt{eqpair}                  & $1.279\pm0.005$        & $1.17^{+0.15}_{-0.10}$ & $1.02\pm0.02$          & $1.065^{+0.000}_{-0.019}$ & $1.088\pm0.017$  & $1.191^{+0.006}_{-0.007}$ & $1.15\pm0.15$ \\
  \raisebox{-1.2ex}{\rule{0pt}{4ex}} \texttt{simpl(kerrbb)}           & $1.278\pm0.005$        & $1.17^{+0.11}_{-0.10}$ & $1.01^{+0.03}_{-0.02}$ & $1.085^{+0.018}_{-0.016}$ & $1.088\pm0.017$  & $1.187^{+0.014}_{-0.012}$ & $1.15\pm0.14$ \\
  \raisebox{-1.2ex}{\rule{0pt}{4ex}} \texttt{simpl(diskbb)}           & $1.288\pm0.016^\dagger$& $1.14^{+0.15}_{-0.11}$ & $0.97\pm0.02$          & $1.009^{+0.018}_{-0.017}$ & $1.038\pm0.017$  & $1.133^{+0.005}_{-0.004}$ & $1.10\pm0.18$ \\
  \hline
 \end{tabular}\\}

 \textbf{Notes.}\\
 \mbox{Quoted errors are statistical uncertainties at the 90\,\% confidence level for the observations, but semi-amplitudes for the sine fits.}\\
 \mbox{$^*$~Note that the \texttt{diskbb\,+\,powerlaw} model overestimates $N_\mathrm{H}$ the more, the more the \texttt{powerlaw} contributes at low energies, see text.}\\
 \mbox{$^\dagger$~As the lack of data above 2\,keV did not allow to constrain the power law with the \texttt{simpl} model, we here used \texttt{diskbb} only.}
\end{table*}
\section{Analysis}\label{sec:analysis}
\begin{figure*}\centering
 \sidecaption%
 \includegraphics[width=12cm]{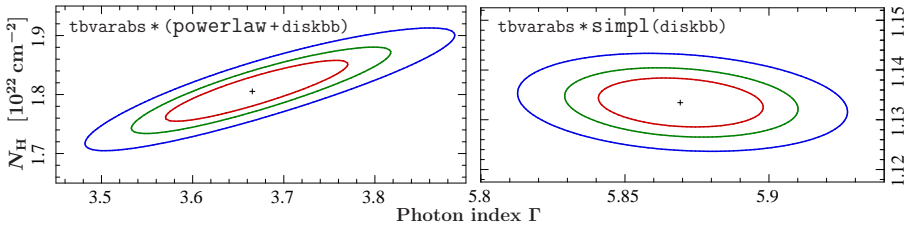}
 \caption{%
  Correlation of the column density $N_\mathrm{H}$ and the photon index $\Gamma$,
  derived with the \texttt{powerlaw} model (left) and the \texttt{simpl} model (right)
  for observation X1.
  The contours show the 68\,\%, 90\,\% and 99\,\% confidence regions
  for two parameters of interest (i.e., $\Delta\chi^2=2.30$, 4.61, and 9.21).
  \vskip1.2\baselineskip
 }
 \label{fig:Gamma-NH_corrrelation}
 \vskip-\baselineskip
\end{figure*}
An overview of previous $N_\mathrm{H}$ measurements for \mbox{LMC\,\,X-1} is given by \citet[Table~2]{Orosz2009}.
We caution, however, that only $<$12\,\% of the hydrogen column density towards the LMC,
$N_\mathrm{H}{=}4\!\times\!10^{21}\,\rm cm^{-2}$ \citep[measured in the LAB 21\,cm survey;][]{Kalberla2005,Bajaja2005},
is of Galactic origin,\footnote{%
 See \url{http://www.astro.uni-bonn.de/~webaiub/english/tools_labsearch.php?alpha=05+39+38.7&beta=-69+44+36}\,.
} while the largest part is detected at $v_{\rm LSR}{=}$200--300\,km\,s$^{-1}$
and thus is likely local to the LMC \citep{Richter1987}.
As the absorption in the 0.5--10\,keV band is mostly caused by metals \citep{Wilms2000},
and the LMC has a much lower metallicity than our Galaxy,
we compile both abundance sets in Table~\ref{tab:LMCabundances}.
The LMC abundances are henceforth used throughout our analysis.

As for all previous observations (see \S\,\ref{sec:intro}),
the X-ray spectra of \mbox{LMC\,\,X-1} investigated here are very soft (see Fig.~\ref{fig:fluxspectra}),
but a hard (albeit very steep) component in addition to a thermal one
is nonetheless needed to describe the data, except for S2 and S3.
The \texttt{powerlaw} model, however, becomes unphysically strong at low energies \citep[e.g.,][]{Shrader1998,DoneZyckiSmith2002}.
A steep photon index $\Gamma{\gg}2$ (e.g., $\Gamma{=}3.7{\pm}0.1$ as measured for X1,
which has the best high-energy coverage due to the EPIC-pn spectrum)
is compensated in spectral fits by an incorrectly strong absorption \citep[e.g.,][]{Yao2005,Suchy2008,Gou2009}.
In contrast, the empirical convolution model \texttt{simpl}
\citep{Steiner2009}
has an intrinsic \emph{low-energy} cut-off
when convolving an input spectrum modeled by, e.g., \texttt{diskbb} or \texttt{kerrbb}.\footnote{%
 Note that -- as a convolution model that relies upon a spectral model
 outside of the energy range spanned by the noticed data --
 \texttt{simpl} must be evaluated on a suitably extended grid.
}
Figure~\ref{fig:Gamma-NH_corrrelation} shows that the
(well known) correlation between $N_\mathrm{H}$ and $\Gamma$
vanishes when \texttt{simpl} is used instead of \texttt{powerlaw}.
Although an even steeper photon index was found using \texttt{simpl},
the value of $N_\mathrm{H}$ is smaller and is more narrowly constrained.

\begin{figure}\centering
 \includegraphics[width=\columnwidth]{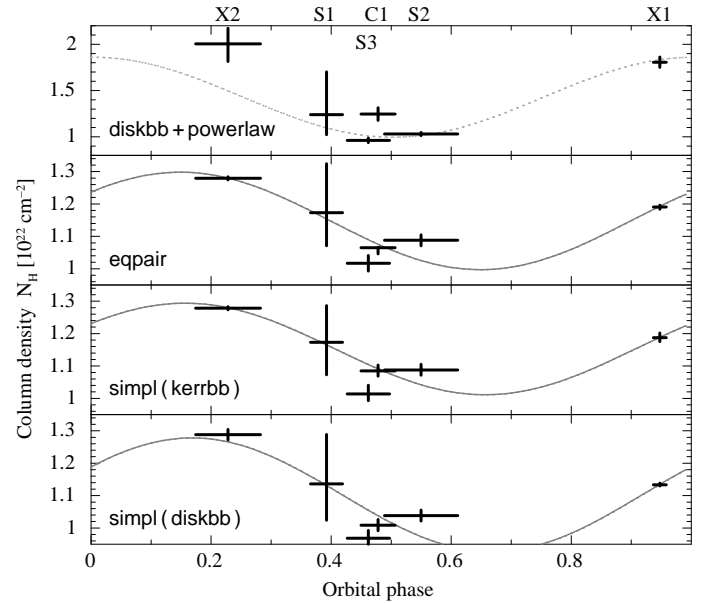}
 \caption{
  $N_\mathrm{H}$ as a function of orbital phase $\phi_\mathrm{orb}^{\rm (T3)}$ (see Table~\ref{tab:obs})
  using various continuum models.
  Note the different scale for the \texttt{diskbb\,+\,powerlaw} model,
  which may predict an unreliably large $N_\mathrm{H}$ (see text).
  The gray lines fit the results with sine curves.
 }
 \label{fig:NH_vs_phi}
 \vskip-1.5\baselineskip
\end{figure}

As the derived absorption might depend on the shape of the continuum,
we investigate different models, namely empirical ones -- such as 
\texttt{diskbb\,+\,powerlaw}, \texttt{simpl(diskbb)},
and \texttt{simpl(kerrbb)} \citep{Li2005} -- as well as the physical Comptonization model
\texttt{eqpair} \citep{Coppi2000}.\footnote{%
  For X1, the $N_\mathrm{H}$ derived with \texttt{diskbb\,+\,compTT} \citep{Titarchuk1994}
  is also consistent with the one from, e.g., \texttt{simpl(diskbb)}.}
Here, these models typically describe the data equally well.
In all fits, the disk has a temperature between 0.65 and 1.1\,keV.
The other parameters, too, are similar to previously obtained values.
Table~\ref{tab:NH} and Fig.~\ref{fig:NH_vs_phi} show our results for the column density
(assuming the LMC abundances given in Table~\ref{tab:LMCabundances})
as a function of orbital phase $\phi_\mathrm{orb}$
for each of the six observations and all four aforementioned continuum models.
In all cases where a steep power law substantially contributes to the model,
the \texttt{diskbb\,+\,powerlaw} model gives
a much higher $N_\mathrm{H}$ than the other models,
due to the systematic error of the \texttt{powerlaw} model.
We therefore ignore these values.
The other models, however, are quite consistent with one another;
their agreement on $N_\mathrm{H}$ is within ${<}8\!\times\!10^{20}\rm\,cm^{-2}$,
which is therefore an upper limit of the systematic error
due to the choice of the continuum.
Using the LMC abundances (Table~\ref{tab:LMCabundances}),
we find column densities in the range of $(1.0{-}1.3)\!\times\!10^{22}\rm\,cm^{-2}$.

We detect a modulation of $N_\mathrm{H}$ with orbital phase:
the observations X1 and X2 close to $\phi_\mathrm{orb}{\approx}0$,
when the BH is behind the donor star,
require a systematically higher $N_\mathrm{H}$ than S3, C1, and S2
close to $\phi_\mathrm{orb}{\approx}0.5$.
In order to quantify this modulation by its mean and amplitude (Table~\ref{tab:NH}),
we fit sine curves to the six measurements for each continuum model (see Fig.~\ref{fig:NH_vs_phi}),
being aware that they do not describe the data very well
and also predict the strongest absorption at $\phi_\mathrm{orb}{=}0.15-0.17$,
which is not expected.

\begin{figure}\centering
 \includegraphics[width=\columnwidth]{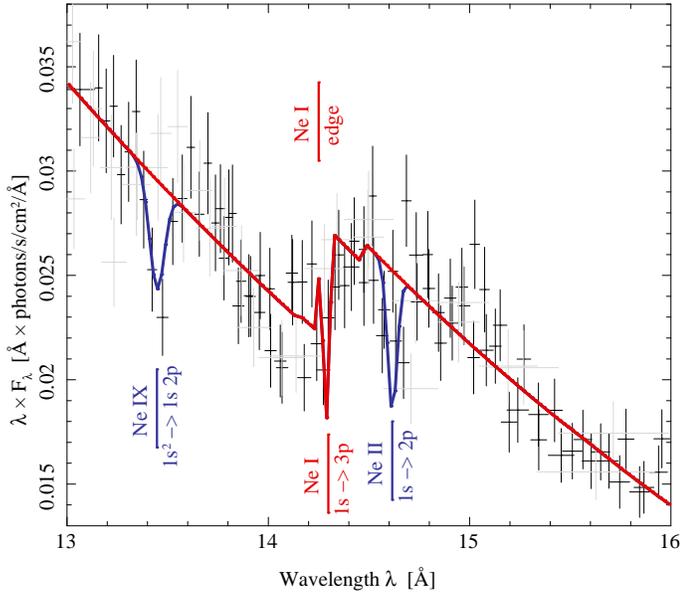}
 \caption{
  The Ne-edge in observation X2.
  The first order RGS spectra (black) reveal
  absorption lines of Ne\,\textsc{ix} at 13.45\,\AA{}
  and probably also Ne\,\textsc{ii} at 13.62\,\AA,
  but the quality of the spectrum does not allow
  for a detailed study of the ionized absorber.
 }
 \label{fig:Needge}
 \vskip-\baselineskip
\end{figure}

Finally, we find marginal evidence for ionized absorption
in the high-resolution spectra (Fig.~\ref{fig:Needge}),
however, a detailed study of these features
is beyond the scope of this paper.

\section{Summary and Discussion}\label{sec:discuss}
The elements with the largest contribution
to the photoabsorption in the soft X-ray band
are significantly less abundant in the LMC than in the Galaxy
(Table~\ref{tab:LMCabundances}).
Because of the lower metallicity,
simply using radio-measured $N_\mathrm{H}$ values in an absorption model
without adopting the LMC abundances
will not allow for a correct description of the physical situation.
Specifically for \mbox{LMC\,\,X-1}, the \emph{equivalent} hydrogen column density
inferred from this X-ray absorption study
-- taking the proper LMC abundances into account --
is actually much higher than the H-column resolved by the LAB survey
\citep[at a half-power beam-width of 0$\fdg$6;][]{Kalberla2005},
which is likely due to additional material in the environment of \mbox{LMC\,\,X-1}
and in the system itself.
This result was not obtained in earlier X-ray absorption measurements,
as erroneously applying Galactic abundances
resulted in smaller $N_\mathrm{H}$ values.

In addition, we have presented the first evidence
that the column density varies in the range $(1.0{-}1.3)\!\times\!10^{22}\rm\,cm^{-2}$.
A modulation with orbital phase
is strongly suggested and would be consistent
with absorption in the stellar wind of the donor giant.
\citet{Orosz2009} assume that the orbital modulation of the X-ray flux
is mostly caused by Thomson scattering in the stellar wind
as they find similar amplitudes\footnote{The fractional full amplitude is here $A=\rm(max{-}min)/mean$.}
in all the three \textsl{RXTE}-ASM energy bands,
namely
$A_{\rm A(1.5-3\,keV)}{\,=\,}7.2{\pm}1.0\,\%$,
$A_{\rm B(3  -5\,keV)}{\,=\,}7.7{\pm}1.1\,\%$, and
$A_{\rm C(5 -12\,keV)}{\,=\,}3.8{\pm}2.9\,\%$.
From a modulation in $N_\mathrm{H}$ with a full amplitude of $3\!\times\!10^{21}\rm\,cm^{-2}$,
$A_{\rm A}{\,=\,}7.7{-}6.9\,\%$,
$A_{\rm B}{\,=\,}1.6{-}2.7\,\%$,
and $A_{\rm C}{\,=\,}0.4{-}1.7\,\%$
are expected,
depending on the assumptions about the ASM response --
i.e., the variation seen with the ASM is almost consistent
with the suggested neutral absorption.
The phase of the current sine fit, however, is not.
Further soft X-ray observations covering more phases are clearly needed,
as the structure of the stellar wind might be more complex than a sine curve.
With three 50\,ks \textsl{Chandra} observations that we have gained for AO\,11,
we will be able to better constrain the modulation.

\acknowledgements
We thank N.\,Przybilla and M.F.\,Nieva for providing the LMC abundances.
M.\,H. and J.\,W. acknowledge funding from the \textsl{Bundesministerium f\"ur Wirtschaft und Technologie}
through the \textsl{Deutsches Zentrum f\"ur Luft- und Raumfahrt} under contract 50OR0701.
We thank the MIT Kavli Institute and the ISSI (Bern)
for their hospitality during the preparation of this work.

\vskip-1.5\baselineskip~

\end{document}